\documentclass[11pt]{article}
\usepackage{amsmath,amstext,amsbsy,amssymb,amsthm,amscd,array}
\usepackage{bm}
\usepackage{graphics,xcolor}
\usepackage{slashed}
\usepackage{cite}
\usepackage{color}
\usepackage{dsfont}
\usepackage{soul} 
\usepackage[utf8]{inputenc}
\newcommand{\half}{{\frac{1}{2}}}
\def\2{{\half}}

\def\bA{{\bm{A}}}

\def\bnabla{\mbox{\boldmath$\nabla$}}

\def\bE{{\bm{E}}}
\def\bB{{\bm{B}}}

\def\bnabla{{\bm{\nabla}}}

\def\bx{{\bm{x}}}

\def\bC{{\bm{C}}}
\def\bZ{{\bm{Z}}}

\def\beq{\begin{equation}}
\def\eeq{\end{equation}}
\def\beqa{\begin{eqnarray}}
\def\eeqa{\end{eqnarray}}

\def\barray{\left(\begin{array}}
\def\earray{\end{array}\right)}
\def\barraynb{\begin{array}}
\def\earraynb{\end{array}}

\def\smallover#1/#2{\hbox{$\textstyle\frac{#1}{#2}$}} %

\begin{document}


\begin{center}

{\Large \bf   
Optical helicity and Hertz vectors}

\vspace{2cm}

\ Mahmut Elbistan

\vspace{5mm}

{\em {\it Institute of Modern Physics, Chinese Academy of Sciences, Lanzhou, China }}\footnote{{\it E-mail address:} elbistan@impcas.ac.cn }

\end{center}

\vspace{3cm}

We study the conserved quantity associated with the dual symmetry of the Maxwell equations,  called the optical helicity, by means of transverse Hertz vectors. In the presence of charges, its evolution yields the integral of $\bm{E}\cdot\bm{B}$ which is the anomaly term for chiral fermions. We also discuss the helicity change in condensed matter systems where topological magnetoelectric effect emerges. An alternative expression of the optical helicity is also found.  Lastly, a dual symmetric Hertz Lagrangian is constructed and its conserved charge is derived.

\vspace{11cm}

\pagebreak



\section{Introduction}

It has long been known \cite{heaviside, larmor} that the vacuum Maxwell equations are invariant under the duality transformations
\beq
\label{tdual}
\bE'=\bE\cos\theta+\bB\sin\theta,\quad
\bB'=\bB\cos\theta-\bE\sin\theta.
\eeq
The conserved charge of this symmetry was obtained by Calkin \cite{calkin} through the Noether theorem as
\beq
\chi=\frac{1}{2}\int d^3\bx\ (\bB\cdot\partial_t \bZ_C-\bZ_C\cdot\partial_t\bB),
\label{Qcalkin}
\eeq
where $\partial_t=\frac{\partial}{\partial t}$. Here $\bm{Z}_C$ is a transverse vector, $\bnabla\cdot\bZ_C=0$, satisfying
\beq
\label{introZ1}
\quad \partial_t \bZ_C=\bA^T, \quad \bnabla\times\bnabla\times\bZ_C=\bE. 
\eeq
$\bA^T$ is the transverse piece of the vector potential. Calkin called $\bZ_C$ the \emph{Hertz vector} \cite{hertz, righi, nisbet, jackson}.

The physical meaning of (\ref{Qcalkin}) becomes clear when it is quantized in a box\cite{calkin}
\beq
\label{pmeanchi}
\hat{\chi}=\sum_{\bm{k}} \hbar (\hat{n}_R(\bm{k}) -\hat{n}_L(\bm{k})).
\eeq
Above, $\hat{n}_{R,L} (\bm{k})$ are the number operators associated with right/left circularly polarized photons for each k-mode, respectively. 
Thus, $\hat{\chi}$ is the operator that corresponds to the difference between the total number of right and left circularly polarized photons.

The same problem was also studied in \cite{DeTe} and the following non-local charge 
\begin{eqnarray}
\chi= \frac{1}{2}\int d^3\bx\  (\bA\cdot\bnabla\times\bA-\bE\cdot\bnabla^{-2}\bnabla\times\bE), 
\label{DTchi}
\end{eqnarray}
was found within the Hamiltonian framework.

Without referring to duality, a covariant generalization of Calkin's conserved current was given by Afanasiev and Stepanovsky \cite{as}. In addition to the usual scalar $A^0$ and  vector $\bm{A}$ potentials,  they introduced dual potentials $C^0(\bx, t)$ and $\bC(\bx, t)$ such that 
\begin{subequations}
\label{ASdefac}
\begin{align}
\bE&=-\bnabla A^0-\partial_t\bA=-\bnabla\times\bC,\\
\bB&=-\bnabla C^0-\partial_t\bC=\bnabla\times\bA.
\end{align}
\end{subequations}  
Their conserved charge is a combination of two Chern-Simons terms
\beq
\chi=\frac{1}{2}\int d^3x\  (\bm{A}\cdot\bB-\bm{C}\cdot\bE).
\label{ashelicity}
\eeq

Later, in \cite{cameronnjp, bliokhnjp}, it was shown that a dual symmetric Lagrangian which includes both potentials $\bA$ and $\bC$ relates this charge (\ref{ashelicity}) to the duality transformations.
The same problem is handled in \cite{EDHZ-heli} within a symplectic approach. At present, $\chi$ (\ref{ashelicity}) is called the \emph{optical helicity} of the electromagnetic field \cite{as,CMYnjp14, bliokhnjp}. For a discussion on the relation between the optical helicity and the photon helicity, see the recent paper \cite{BBr2}.

The duality symmetry of the Maxwell equations were also studied in \cite{hb} by means of the conventional Hertz vectors (polarization potentials) $\bm{\Pi}_e$ and $\bm{\Pi}_m$ \cite{jackson}. By using these Hertz vectors, the gauge potentials in \cite{hb} are defined as
\begin{subequations}
\label{chertz}
\begin{align}
\bA&=\partial_t\bm{\Pi}_e+\bnabla\times\bm{\Pi}_m, \quad A^0=-\bnabla\cdot\bm{\Pi}_e,\\
\bC&=\partial_t\bm{\Pi}_m-\bnabla\times\bm{\Pi}_e, \quad C^0=-\bnabla\cdot\bm{\Pi}_m,
\end{align}
\end{subequations}
where the Lorenz gauge condition is already built in \cite{jackson}. However, the conserved charge (\ref{ashelicity}) was not discussed there.
The Hertz vectors were also revisited in \cite{BBr1, bliokhst} in a complex form.
 
Recently, the concepts of duality and helicity become a useful tool for the investigation of light-matter interactions in various materials\cite{fcorbaton}, including chiral and magnetoelectric molecules \cite{bliokhm,bliokhcm}. 
In this work, we study the duality symmetry and the optical helicity by using a different set of Hertz vectors. In particular, we discuss the evolution of the optical helicity (\ref{ashelicity}) in physical systems like Weyl semimetals \cite{zb} and topological insulators \cite{qhz} which exhibit the topological magnetoelectric effect.


\section{Duality and the optical helicity}
\label{basics}

In the vacuum not only the magnetic field, but also the electric field becomes transverse, i.e., $\bE=\bE^T$ and $\bB=\bB^T$. We can write them by means of the gauge invariant transverse potentials as
$$
\bE=-\partial_t\bA^T=-\bnabla\times\bC^T,\quad
\bB=-\partial_t\bC^T=\bnabla\times\bA^T,
$$
cf (\ref{ASdefac}).
Therefore, $\chi$ (\ref{ashelicity}) can be expressed only in terms of $\bB^T, \bE^T, \bA^T$ and $\bC^T$ such that
\beq
\label{helicityt}
\chi=\frac{1}{2}\int d^3x\  (\bm{A}^T\cdot\bB^T-\bm{C}^T\cdot\bE^T),
\eeq
where we ignore the surface terms. This shows its gauge invariance.
Moreover, it is explicitly dual symmetric since the duality transformations (\ref{tdual}) descend to the transverse potentials as
\beq
\label{tdualACt}
{\bA^T}'=\bA^T\cos\theta+\bC^T\sin\theta,\quad
{\bC^T}'=\bC^T\cos\theta-\bA^T\sin\theta.
\eeq

On the other hand, although it is the conserved charge of the duality symmetry, Calkin's charge (\ref{Qcalkin}) itself is not manifestly dual symmetric. Moreover, since it is defined in terms of a Hertz vector, its gauge invariance is not explicit \cite{nisbet}.
Even so, by combining (\ref{introZ1}) and (\ref{ASdefac}), we obtain
 \beq
\label{Z1ACt}
\bnabla\times\bZ_C=-\bm{C}^T,\quad \partial_t \bZ_C=\bA^T,
\eeq   
which shows that the two optical helicity expressions, (\ref{Qcalkin}) and (\ref{ashelicity}) are indeed equivalent.
 
We would like to emphasize an important fact. Since
$\bm{A}$ alone is sufficient to define $\bE$ and $\bB$, the second vector potential $\bm{C}$ seems to be superfluous in (\ref{ASdefac}). However, the advantage of using $\bC$ in (\ref{ashelicity}) is to obtain an explicitly gauge and dual invariant helicity expression. 

The optical helicity $\chi$ (\ref{ashelicity}) is composed of a magnetic part $\chi_m=\int d^3\bx\ \bA\cdot\bB$ and an electric part $ \chi_e=\int d^3\bx\ \bC\cdot\bE$ \footnote{Equivalently, one can redefine the magnetic part as $\chi_m'=\int d^3\bx\ \bA\cdot\bnabla\times\bA$ and the electric part as $ \chi_e'=\int d^3\bx\ \bC\cdot\bnabla\times\bC$,
such that the total helicity becomes a sum, i.e., 
$
\chi=\frac{1}{2}(\chi_m'+\chi_e').
$}.
In the vacuum, each one evolves as
\beq
\label{chievo}
\frac{d \chi_m}{dt}=-2\int d^3\bx\ \bE\cdot\bB, \quad
\frac{d \chi_e}{dt}=-2\int d^3\bx\ \bE\cdot\bB.
\eeq
$\chi_m$ is known as the \emph{magnetic helicity}. It is a degree of the knottedness of the magnetic field when the Pfaffian of the electromagnetic field strength vanishes i.e.  $\bE\cdot\bB=0$ \cite{ranada2, knotsrev}.
The first equality in (\ref{chievo}) remains valid in the presence of electric sources and it relates $\chi_m$ and the chiral anomaly of massless fermions \cite{giosha}. 

When the electric $4$-current $j^\mu=(j^0, \bm {j})$ is switched on, 
 \begin{subequations}
\label{MaxwellEBes}
\begin{align}
\label{Maxwellsourcees}
\bm{\nabla}\cdot\bE=j^0(\bx, t), &\quad \bm{\nabla}\times\bB-\frac{\partial \bE}{\partial t}=\bm{j}(\bx, t),\\
\bm{\nabla}\cdot\bB=0, &\quad \bm{\nabla}\times\bE+\frac{\partial \bB}{\partial t}=0,
\label{Maxwellfreesourcees}
\end{align}
\end{subequations}  
the duality symmetry is broken and $\chi$ is no longer conserved. However, it retains its meaning even in this situation. Being defined in terms of the  transverse fields, it is the generator of the duality transformations of $\bE^T,\bB^T$ (\ref{tdual}) and $\bA^T, \bC^T$ (\ref{tdualACt}). In addition, its quantization still results in the sum over the helicities of individual plane-wave modes\footnote{Calculation in \cite{rcam} was done in Schroedinger picture.} \cite{rcam}, thus $\hat{\chi}$ still yields (\ref{pmeanchi}). 

We would like to study the optical helicity (\ref{ashelicity}) and its evolution by means of the Hertz vectors. As seen from (\ref{helicityt}) in the vacuum $\chi$ is defined in terms of the transverse vectors. In the presence of charges and polarizations, one can always adopt Coulomb gauge and continue to deal with the transverse fields. However, the conventional Hertz vectors $\bm{\Pi}_e$ and $\bm{\Pi}_m$  are built within Lorenz gauge. Thus, they are not suitable for our aim. Moreover,  
they are originally invented to deal with the electric and magnetic polarizations. Therefore, the expressions of the electromagnetic fields and the potentials (\ref{chertz}) in terms of them are complicated. We need simpler descriptions.

In fact, such a description is already provided with $\bZ_C$ (\ref{introZ1}). Since electromagnetic theory possesses two Hertz vectors linked by duality, we shall introduce another Hertz vector which is dual to $\bZ_C$ and which yields $\bC^T$ and $\bB$ in a simple manner.

In the next section, we follow the route outlined in \cite{calkin} and derive a dual Hertz vector which naturally emerges from the transformations of $\bm{C}$. Therefore, we will obtain a set of transverse Hertz vectors which is different from (\ref{chertz}). 


\section{Transverse Hertz vectors} 

\subsection{The dual Hertz vector}

Motivated by the presence of the dual fields (\ref{tdual}) and the potentials (\ref{tdualACt}), we look  for a dual Hertz vector.
To get it, we modify the arguments in \cite{calkin} by defining $\bE$ and $\bB$ in terms of dual potentials (\ref{ASdefac}).

With  $C^0$ and $\bm{C}$, the  infinitesimal version of (\ref{tdual}) reads 
\begin{eqnarray*}
\bnabla\times\delta\bm{C}=\delta\theta (\partial_t \bm{C}+\bnabla C^0),\quad
-\partial_t \delta\bm{C}-\bnabla \delta C^0=\delta \theta (\bnabla\times\bm{C}).
\end{eqnarray*}
We propose the solutions
$$
\delta C^0=-\delta\theta (\partial_t \lambda),\quad \delta\bm{C}=\delta\theta( \bnabla\lambda-\bnabla\times\bZ_A), 
$$
where $\lambda$ is arbitrary and $\bZ_A$ is the {\it{second Hertz vector}} which satisfies
\begin{subequations}
\label{Z2EBt}
\begin{align}
\bnabla\times (\bnabla\times\bZ_A)&=-\bnabla C^0-\partial_t\bm{C}\equiv\bB,\\
\partial_t(\bnabla\times\bZ_A)&=\bnabla\times\bC\equiv-\bE.
\end{align}
\end{subequations}
We deduce 
$$
\bnabla\times\bZ_A=\bA^T,\quad
\partial_t \bZ_A=\bC-\bnabla\zeta_A,
$$
where $\zeta_A(t,\bx) $ is a gauge parameter. 
This arbitrariness can, as in (\ref{introZ1}), be removed by choosing
$\bnabla\cdot\bZ_A=0$, i.e., $\bZ_A\equiv\bZ_A^T$
yielding
\begin{eqnarray}
\label{Z2ACt}
\bnabla\times\bZ_A=\bA^T, \quad \partial_t \bZ_A=\bC^T.
\end{eqnarray} 
We obtain the following Noether charge from the variation of the free Maxwell action
\beq
\label{Qcalkindual}
\chi=\frac{1}{2}\int d^3\bx\ (\bZ_A\cdot\partial_t\bE-\bE\cdot\partial_t \bZ_A),
\eeq
cf. (\ref{Qcalkin}).
From (\ref{Z1ACt}) and (\ref{Z2ACt}), we can derive integral solutions for the transverse Hertz vectors
\begin{subequations}
\label{ZH}
\begin{align}
\label {ZAB}
\bm{Z}_A(\bx, t)&=\int d^3\bx'\ D(\bx-\bx')\bB^T(\bx', t),\\
\bm{Z}_C(\bx, t)&=\int d^3\bx'\ D(\bx-\bx')\bE^T(\bx', t),
\label{ZCE}
\end{align}
\end{subequations}
with the help of the Helmholtz theorem. Here, $D(\bx-\bx')$ is the Green's function for the 3-dimensional Laplacian $\bnabla^2$
$$
 D(\bx-\bx')=\frac{1}{4\pi|\bx-\bx'|}=\int \frac{d^3\bm{k}}{(2\pi)^3}\ \frac{e^{-i\bm{k}\cdot(\bx-\bx')}}{|\bm{k}|^2}, \quad -\bnabla^2D(\bx-\bx')=\delta(\bx-\bx').
$$
When the sources are switched off in (\ref{MaxwellEBes}), we can safely write $\bB^T=\bB,\ \bE^T=\bE$ in (\ref{ZH}). The Hertz vectors are originally subject to gauge transformations \cite{nisbet}. However, our $\bZ_{A,C}$ (\ref{ZH}) are gauge fixed and unique. 

Moreover, by using (\ref{ZH}) and (\ref{tdual}), we find that they are dual to each other:
\beq
\label{zdualAC}
{\bZ_A}'= \bZ_A\cos\theta -\bZ_C\sin\theta,\quad
{\bZ_C}'= \bZ_C\cos\theta +\bZ_A\sin\theta.
\eeq

Hereafter, we consider (\ref{ZH}) as the definitions of our Hertz vectors.

Alternatively, we can also derive $\bZ_A$ by following the arguments in \cite{jackson} but choosing Coulomb gauge instead of Lorenz gauge.

Lastly, we would like to note that in \cite{bliokhst} (see also \cite{BBr1}), $\bE$ and $\bB$ were defined by means of the conventional Hertz vectors (\ref{chertz}) as
$$
\bm{E}=\bnabla\times\bnabla\times\bm{\Pi}_e-\bnabla\times\partial_t\bm{\Pi}_m,\quad \bm{B}=\bnabla\times\bnabla\times\bm{\Pi}_m+\bnabla\times\partial_t\bm{\Pi}_e.
$$
However, their definition of the transverse potentials, 
$$
\bA^T=\bnabla\times\bm{\Pi}_m,\quad \bC^T=-\bnabla\times\bm{\Pi}_e,
$$
matches with our (\ref{Z1ACt}) and (\ref{Z2ACt}).


\subsection{Properties of the Hertz vectors}

\subsubsection{Free theory and relation to the Maxwell equations}
\label{RelationMaxwell}

We would like to show the similarity between the Maxwell equations and the equations of the Hertz vectors (\ref{Z1ACt}) and (\ref{Z2ACt}).

Using the expressions (\ref{ZH}), we find   
$$
\bnabla\times\bZ_A-\partial_t\bZ_C=\int d^3\bx'\ D(\bx-\bx') \Big( \bnabla'\times\bB(\bx', t)-\partial_t\bE(\bx', t) \Big)=0.
$$
A similar relation holds for the other equations
$$
\bnabla\times\bE+\partial_t\bB=0 \implies \bnabla\times\bZ_C+\partial_t\bZ_A=0.
$$
Together with the transversality  of (\ref{ZH}), 
the Hertz vectors provide a set of Maxwell-like equations
\begin{subequations}
\label{HertzMaxwell}
\begin{align}
\label{hertzsource}
\bm{\nabla}\cdot\bZ_C=0, &\quad \bm{\nabla}\times\bZ_A-\frac{\partial \bZ_C}{\partial t}=0,\\
\bm{\nabla}\cdot\bZ_A=0, &\quad \bm{\nabla}\times\bZ_C+\frac{\partial \bZ_A}{\partial t}=0,
\label{hertzfreesource}
\end{align}
\end{subequations}
which are in one-to-one correspondence with the vacuum Maxwell equations. 

From (\ref{HertzMaxwell}), we observe that $\bZ_{A, C}$ satisfy the homogenous wave equations 
\beq
\label{Hertzwave}
\Box\bZ_A=0,\quad \Box\bZ_C=0,\quad    \Box\equiv\partial_t^2-\bnabla^2,
\eeq
like usual Hertz vectors (\ref{chertz}) do when the polarizations are switched off \cite{jackson}. On the other hand, free fields can be written in terms of  $\bZ_{A, C}$ in a simpler way
\begin{subequations}
\label{hertzdefac}
\begin{align}
\bE&=\bnabla\times\bnabla\times\bZ_C=-\partial_t^2\bZ_C,\\
\bB&=\bnabla\times\bnabla\times\bZ_A=-\partial_t^2\bZ_A.
\end{align}
\end{subequations}
We observe that $\bZ_A$ and $\bZ_C$ become potentials for the gauge fields $\bm{A}$ and $\bm{C}$.

In a Lorentz covariant form the Maxwell equations read  
$
\partial_\mu F^{\mu\nu}=0, \, \partial_\mu \tilde{F}^{\mu\nu}=0, 
$ 
where $F^{0i}=-E^i, \ F^{ij}=-\epsilon^{ijk}B^k$, 
$\tilde{F}^{0i}=-B^i, \ \tilde{F}^{ijk}=\epsilon^{ijk}E^k$.
Since the Hertz vectors satisfy similar equations (\ref{HertzMaxwell}), we are tempted to define a quantity $H_{\mu\nu}$\footnote{A similar but Lorentz covariant tensor is
considered in \cite{hb}.} such that
$$
H^{0i}=-Z^i_C,\quad H^{ij}=-\epsilon^{ijk}Z^k_A.
$$
Then (\ref{HertzMaxwell}) can be written in a compact form as 
\beq
\label{HertzMaxwell2}
\partial_\mu H^{\mu\nu}=0, \quad \partial_\mu \tilde{H}^{\mu\nu}=0,
\eeq
where $\tilde{H}^{\mu\nu}=\frac{1}{2}\epsilon^{\mu\nu\rho\sigma}H_{\rho\sigma}$ is the dual  of $H^{\mu\nu}$ i.e., 
$
\tilde{H}^{0i}=-Z^i_A, \  \tilde{H}^{ij}=\epsilon^{ijk}Z^k_C
$.
We note that (\ref{HertzMaxwell2}) is not a Lorentz covariant equation because our Hertz vectors only generate the transverse parts of the electromagnetic potentials:
\beq
\label{ZACt2}
-\partial_0 H^{0i}=-\partial_j H^{ij}=A^i ,\quad  -\partial_0 \tilde{H}^{0i}=-\partial_j \tilde{H}^{ij}=C^i.
\eeq
Equations (\ref{Hertzwave}) can be combined to give a single equation,
$
\Box H^{\mu\nu}=0.
$
The fields 
$F^{\mu\nu}$ and $H^{\mu\nu}$ are related 
as
\beq
\label{tenrl}
F^{\mu\nu}=-\bnabla^2 H^{\mu\nu},
\eeq 
which is the compact form of (\ref{ZH}). Likewise, the duality transformations (\ref{zdualAC}) can be expressed as 
$
H'^{\mu\nu}=H^{\mu\nu}\cos\theta+{\tilde{H}}^{\mu\nu}\sin\theta.
$ 


\subsubsection{Turning on polarizations}

After turning on the electric $\bm{P}(\bx, t)$ and the magnetic $\bm{M}(\bx, t)$  polarizations,
Maxwell's equations become
\begin{subequations}
\label{MaxwellEBps}
\begin{align}
\bm{\nabla}\cdot\bE=-\bnabla\cdot\bm{P}, &\quad \bm{\nabla}\times\bB-\frac{\partial \bE}{\partial t}=\frac{\partial \bm{P}}{\partial t}+\bnabla\times\bm{M},\\
\bm{\nabla}\cdot\bB=0,\quad &\quad \bm{\nabla}\times\bE+\frac{\partial \bB}{\partial t}=0.
\end{align}
\end{subequations}  
They can be written as
$
\partial_\mu  F^{\mu\nu}=l^\nu, \ \partial_\mu \tilde{F}^{\mu\nu}=0, 
$
where 
\beq
\label{defell}
l^0=-\bnabla\cdot\bm{P}, \quad \bm{l}=\frac{\partial \bm{P}}{\partial t}+\bnabla\times\bm{M}, \quad \partial_\mu l^\mu=0.
\eeq

Recalling the definitions (\ref{ZH}), we observe that all the Maxwell-like equations (\ref{HertzMaxwell}) are valid except the modified one below
\beq
\label{HertzMaxwellps}
\bm{\nabla}\times\bZ_A-\frac{\partial \bZ_C}{\partial t}= \int d^3\bx'\ D(\bx-\bx')\ \bm{l}^T(\bx', t),
\eeq
where $\bm{l}^T=\partial_t \bm{P}^T+\bnabla\times\bm{M}^T$. In addition, the wave equations (\ref{Hertzwave}) become inhomogenous
\begin{subequations}
\label{waveHps}
\begin{align}
\Big( \frac{\partial^2}{\partial t^2}-\bnabla^2\Big) \bZ_A&=\int d^3\bx'\ D(\bx-\bx')\bnabla'\times\bm{l}^T,\\
\Big( \frac{\partial^2}{\partial t^2}-\bnabla^2\Big) \bZ_C&=-\int d^3\bx'\ D(\bx-\bx')\frac{\partial \bm{l}^T}{\partial t},
\end{align}
\end{subequations}
Simple formulas like (\ref{ZACt2}) and (\ref{tenrl}) are still valid for the transverse vectors. 

We emphasize that the equations (\ref{HertzMaxwellps}) and (\ref{waveHps}) of $\bZ_{A, C}$ are different from the ones of $\bm{\Pi}_{e,m}$ \cite{jackson}.


\section{Applications of the Hertz vectors}

\subsection{Evolution of the optical helicity}

Let us consider the evolution of $\chi$ in the presence of sources (\ref{MaxwellEBes}). In the Coulomb gauge, we have $\bA=\bA^T,\ \bC=\bC^T,\ C^0=0$ and most of the previous definitions are the same (see also \cite{rcam}), 
\begin{eqnarray*}
\bB=\bnabla\times\bA^T,\quad \bE^L=-\bnabla A^0,\quad \bE^T=-\bnabla\times\bC^T=-\partial_t\bA^T,
\end{eqnarray*}
except the one below,
$$
\partial_t \bC^T(\bx, t)=\int d^3\bx'\ D(\bx-\bx') \bnabla'\times\bm{j}^T(\bx', t)-\bB(\bx, t).
$$
The Maxwell-like equations satisfied by the Hertz vectors (\ref{HertzMaxwell}) are also modified
\begin{subequations}
\label{HertzMaxwelles}
\begin{align}
\label{hertzsourcees}
\quad \bm{\nabla}\times\bZ_A-\frac{\partial \bZ_C}{\partial t}&=\int d^3\bx'\ D(\bx-\bx') \ \bm{j}^T(\bx', t) ,\\
\quad \bm{\nabla}\times\bZ_C+\frac{\partial \bZ_A}{\partial t}&=0.
\label{hertzfreesourcees}
\end{align}
\end{subequations}
They have the same form as (\ref{HertzMaxwellps}).
The optical helicity 
\beq
\label{helielectric}
\chi = \frac{1}{2}\int d^3\bx\  (\bA^T\cdot\bB-\bm{C}^T\cdot\bE^T),
\eeq
evolves as
\beq
\label{hevoes}
\frac{d\chi}{dt}=\int d^3\bx\  \bm{j}\cdot\bm{C}^T.
\eeq
This was obtained before in \cite{rcam}. In order to go further, we recall our transverse Hertz vectors $\bZ_{A,C}$. Usage of either (\ref{Z1ACt}) or (\ref{HertzMaxwelles}) together with  (\ref{MaxwellEBes}) allows us to rewrite (\ref{hevoes}) as 
\begin{eqnarray}
\frac{d\chi}{dt}&=&-\int d^3\bx\ \bE\cdot\bB-\int  d^3\bx\ d^3\bx'\ D(\bx-\bx')\bE(\bx', t)\cdot\partial_t^2\bB(\bx, t),
\label{tdchiresult}
\end{eqnarray}
which relates the optical helicity of the electromagnetic field to the anomaly term $\bE\cdot\bB$. Therefore, in a suitable system, (\ref{tdchiresult}) may determine the interplay between the optical helicity and the anomaly. We emphasize that our result is different from the individual equations of $\chi_{m}$ and $\chi_e$ in (\ref{chievo}). The right hand side of (\ref{tdchiresult}) vanishes when $j^\mu$ is switched off, as it should.

Let us consider the condensed matter systems for physical examples of the optical helicity evolution. We will focus on physical systems where anomaly related transport phenomena is realized. Such systems may exhibit \emph{effective polarizations}. 
Therefore, it is important to emphasize that (\ref{tdchiresult}) is still valid when $j^\mu$ is replaced by the polarization vector $l^\mu$ (\ref{defell}) in calculations.

Firstly, we consider a $3-d$ condensed matter system called the Weyl semimetal which exhibits anomaly and anomaly induced effects like the chiral magnetic effect \cite{zb}. In \cite{eweyl}, a Weyl semimetal model which includes $2-$component Weyl spinors and two gauge fields was studied. As a result, the triangle diagram calculation yielded a gauge anomaly and the effective action of the Weyl semimetal becomes\footnote{The well-known low energy model of \cite{zb} was built with $4-$component Dirac bi-spinors and it contains both a vector and an axial current. Integration of the fermions in the path integral yields the chiral anomaly and therefore  the second term in their effective action has the coefficient $\frac{1}{32\pi^2}$. } 
\beq
\label{effaMtheta}
S_{eff}=-\frac{1}{4}\int d^4x\ F^{\mu\nu}F_{\mu\nu}-\frac{1}{48 \pi^2}\int d^4x \ \Theta(\bx, t) F_{\mu\nu}\tilde{F}^{\mu\nu},
\eeq
where the second term comes from 1-loop calculations as a quantum correction. The modified Maxwell equations become
\begin{subequations}
\label{MaxwellWeyl}
\begin{align}
\bm{\nabla}\cdot\bE=\frac{1}{12\pi^2}(\bnabla\Theta)\cdot\bB, \quad &\bm{\nabla}\times\bB-\frac{\partial \bE}{\partial t}=-\frac{1}{12\pi^2}\Big((\partial_t\Theta)\bB+(\bnabla\Theta)\times\bE\Big),\\
\bm{\nabla}\cdot\bB=0,\quad\quad\quad \quad &\bm{\nabla}\times\bE+\frac{\partial \bB}{\partial t}=0. 
\end{align}
\end{subequations}
Above $\Theta(\bx, t)$ is a scalar which is intrinsic to the material \cite{zb}.

Comparing (\ref{MaxwellWeyl}) with (\ref{MaxwellEBps}), we observe that $\bB$ induces an electric polarization, and vice versa. This is the topological magnetoelectric effect \cite{ qhz}, a phenomenon known as the axion electrodynamics in high energy physics \cite{wilczek}. 

Since the derivation of (\ref{tdchiresult}) is general, we replace $j^\mu$ in (\ref{MaxwellEBes}) with the polarization 4-vector $\l^\mu$ which, for our particular case (\ref{MaxwellWeyl}), becomes  
\beq
l^0=\frac{1}{12\pi^2}(\bnabla\Theta)\cdot\bB, \quad \bm{l}=-\frac{1}{12\pi^2}\Big((\partial_t\Theta)\bB+(\bnabla\Theta)\times\bE\Big).
\eeq
From (\ref{MaxwellWeyl}) we find
\beq
\label{waveBpolar}
\partial_t^2\bB=\bnabla^2\bB-\frac{1}{12\pi^2}\bnabla\times\Big((\partial_t\Theta)\bB+(\bnabla\Theta)\times\bE\Big).
\eeq
Substitution of (\ref{waveBpolar}) in (\ref{tdchiresult}) cancels the anomaly term $\int d^3\bx\ \bE\cdot\bB$ and we get a new equation which relates the optical helicity to the properties of the Weyl semimetal
\beq
\label{tdchiresultws}
\frac{d\chi}{dt}=\frac{1}{12\pi^2}\int d^3\bx'\ D(\bx-\bx') \bE(\bx', t)\cdot\bnabla\times\Big((\partial_t\Theta)\bB+(\bnabla\Theta)\times\bE\Big). 
\eeq
This result  (\ref{tdchiresultws}) shows the evolution of the optical helicity in a Weyl semimetal by connecting it to the structure of the material. Thus, it may provide a new perspective, which is based on the duality and the optical helicity, for the analysis of that semimetal. 
  
Indeed, duality and helicity are important tools for the investigation of the interaction of light with matter \cite{fcorbaton}, e.g. the magnetoelectric matter (see \cite{bliokhcm} and the references therein). 

In \cite{hky}, the possibility of a relation between the magnetic helicity $\chi_m$ and the emission of circularly polarized photons in Dirac semimetals is mentioned. Here, instead of $\chi_m$, we consider the evolution of $\chi$ because in the context of the duality it is the meaningful quantity and its quantization naturally yields (\ref{pmeanchi}).

Similar ideas may apply to $3+1$ dimensional topological insulators which also host the topological magnetoelectric effect (\ref{MaxwellWeyl}). They can be described with a similar effective action (\ref{effaMtheta}) and the corresponding equation (\ref{tdchiresultws}) with the change $\Theta(\bx, t)\to P_3(\bx, t)$ where $P_3$ is the magnetoelectric polarization \cite{qhz}. 

Indeed, in \cite{bnp}, a  Berry phase\footnote{The calculation was performed at $t=0$.} was derived through the adiabatic change of the parameters like $\Theta$ from a Lagrangian similar to (\ref{effaMtheta}). This phase accompanies the photon states and interestingly its explicit calculation results in $\hat{\chi}$  (\ref{pmeanchi}). There, it was argued that the effects of this phase might be visible in topological insulators. We conclude that (\ref{tdchiresultws}) is the classical analogue of the evolution of that phase.

Since they are described with equations similar to (\ref{MaxwellWeyl}), the above discussions may apply to the axion crystals \cite{yamaxion}.


\subsection{Dual symmetric version of Calkin's charge}

In Section \ref{basics}, we presented different expressions for the optical helicity which have the same physical meaning.
Calkin's charge (\ref{Qcalkin}) is not explicitly dual symmetric because Calkin used a single Hertz vector $\bm{Z}_C$ (and a single vector potential). But, physical quantities in the free Maxwell theory should be dual symmetric \cite{bcy}. Therefore, it is desirable to put (\ref{Qcalkin}) in a form so that it enjoys this symmetry.

Since we have two Hertz vectors, which are dual to each other (\ref{zdualAC}), we can write a dual symmetric version of  (\ref{Qcalkin}) 
\beq
\chi=\frac{1}{2}\int d^3\bx\ (\bB\cdot\partial_t\bZ_C-\bE\cdot\partial_t\bZ_A),
\label{dualsymQc}
\eeq 
where we combine (\ref{Qcalkin}) and (\ref{Qcalkindual}). This expression is explicitly gauge invariant. One can show that it is equivalent to the other conserved charges, namely (\ref{Qcalkin}), (\ref{DTchi}) and (\ref{ashelicity}). Thus it can be considered an alternative expression for the optical helicity.


\subsection{ A dual symmetric Hertz vector Lagrangian}

In Section \ref{RelationMaxwell}, we found that the Hertz vectors satisfy the wave equations (\ref{Hertzwave}). Such equations can be derived naturally from a Klein-Gordon type action.

Recently, we worked on the dual symmetric Klein-Gordon Lagrangians \cite{EHZdual2} within the photon wave function approach \cite{BBrev}. Here, we would like to adopt the same framework by using the complex Hertz vectors\footnote{A similar complex vector was introduced before in \cite{BBr1, bliokhst} with different Hertz potentials.}
$$
\bm{Z}_\pm\equiv \frac{1}{\sqrt{2}}(\bZ_C\pm i\bZ_A),
$$
which satisfy
\beq
\label{BBHertz}
i\partial_t \bZ_\pm=\mp i (\bm{S}\cdot\bnabla) \bZ_\pm,
\eeq 
where $(S_i)_{ab}=-i\epsilon_{iab}$ are the generators of rotations for spin-1 particles.  We note that (\ref{BBHertz}) is the compact form of (\ref{HertzMaxwell}).  
Duality transformations (\ref{zdualAC}) become
\begin{eqnarray}
\label{dsymZ}
\bm{Z}_\pm'=e^{\mp i\theta}\bm{Z}_\pm.
\end{eqnarray}
We build a Klein-Gordon type (lower-derivative) Lagrangian \footnote{$L_Z$ (\ref{KGLagZ}) does not have correct dimensions. We will only treat it as a mathematical tool which generates correct equations of motion and which makes it easier to investigate the symmetries.}
\beq
\label{KGLagZ}
L_Z=\frac{1}{2}(\partial_\mu \bm{Z}_-)\cdot(\partial^\mu \bm{Z}_+).
\eeq
Variation of (\ref{KGLagZ}) with respect to $\bZ_\pm$ yields (\ref{Hertzwave}). $L_Z$ is dual symmetric and the conserved charge is
\beq
\label{hertzcharge}
Q=\frac{1}{2}\int d^3x\ \Big(\bZ_A\cdot \bA^T-\bZ_C\cdot \bC^T \Big).
\eeq
Similar to $\chi$ (\ref{ashelicity}), $Q$ is in the form of a double Chern-Simons integral. In fact, one can obtain $\chi$ from $Q$ with the substitution  $\bZ_\pm\to(\bA\pm i\bC)$. 

Classically, $Q$ seems to be an acceptable quantity. However, after quantization it yields
\beq
\label{hertzchargelm}
\hat{Q}=\sum_{\bm{k}} \frac{\hbar (\hat{n}_R(\bm{k}) -\hat{n}_L(\bm{k}))}{|\bm{k}|^2},
\eeq
which may suffer an infrared divergence for $|\bm{k}|= 0$. 
We conjecture that, the action of $\hat{Q}$ on the photon states with $|\bm{k}|=0$ can be excluded in a non-interacting theory. Another possibility is that the difference $\frac{\hat{n}_R(\bm{k})}{|\bm{k}|^2}-\frac{\hat{n}_L(\bm{k})}{|\bm{k}|^2}$ may be finite in the limit $|\bm{k}|\to 0$.  

It is easy to obtain Lipkin's 00th zilch as the conserved charge of the duality symmetry with the replacement $\bZ_\pm\to (\bE\pm i\bB)$ in (\ref{KGLagZ}). Moreover, replacing electromagnetic fields with their curls, one can get higher order Lagrangians which are again dual symmetric. As it is noted in \cite{EHZdual2}, one can derive infinitely many conserved charges of free electromagnetism (see \cite{CMYnjp14} and the references therein). In our procedure, with the definitions (\ref{dsymZ}) and (\ref{KGLagZ}), it is clear that those charges are all related to the duality symmetry.


\section{Discussion}

We studied the optical helicity $\chi$ (\ref{ashelicity}) by means of the transverse Hertz vectors. Since our Hertz vectors are different from the conventional ones, we explored their physical properties. Solutions of their Maxwell-like equations were simply given in terms of the electromagnetic fields (\ref{ZH}). Their duality transformations were presented (\ref{zdualAC}). We also investigated their behavior in the presence of polarizations. 

As a first application, we considered the evolution of the optical helicity. We found that, when electric charges are added, there exists a generic relation between the optical helicity and the anomaly term (\ref{tdchiresult}). As physical examples, we considered condensed matter systems like Weyl semimetal and $3+1$-d topological insulator where effective polarizations emerge. We found an interesting equation (\ref{tdchiresultws}) which relates the optical helicity to the effective description of the Weyl semimetal. Our approach based on the helicity and the duality is, of course, preliminary. Yet, from the theoretical side, it may provide a new angle to study the material.  The same arguments also apply to the topological insulators.
For future study, we would like to work out the quantized version of (\ref{tdchiresultws}). 

We also obtained an alternative dual symmetric expression of $\chi$ in terms of the Hertz vectors (\ref{dualsymQc}). 
Lastly, we proposed a dual symmetric Hertz vector Lagrangian and computed its Noether charge $Q$ (\ref{hertzcharge}). This charge may be a new conserved quantity, however its quantization needs further clarification. Our framework is useful to generate the infinite hierarchy of conserved quantities as a result of duality symmetry.


\subsection*{Acknowledgments}

I would like to thank  J. Balog, Ö. F. Dayı, C. J. Halcrow and P. Horvathy for their comments on the manuscript. I am also grateful to the members of our High Energy Nuclear Physics Group at IMP for fruitful discussions. This work was supported by the Chinese Academy of Sciences President's International Fellowship Initiative (No. 2017PM0045).



\begin{thebibliography}{99}

\bibitem{heaviside}
O. Heaviside, {\it{On the forces, stresses and fluxes of energy in the electromagnetic field}}, Phil. Trans. R. Soc. A 183 (1892) 423 

\bibitem{larmor}
 J. Larmor, {\it{A Dynamical theory the electric and luminiferous medium}}, Phil. Trans. R. Soc. A 190 (1897) 205 

\bibitem{calkin} M. G. Calkin, {\it{An invariance property of the free electromagnetic field}}, Am. J. Phys. 33 (1965) 958 

\bibitem{hertz} H. Hertz, {\it{Die Krafte electrischer Schwingungen, behandelt nach der Maxwell’schen Theorie}}, Ann. Phys. 36 (1889) 1

\bibitem{righi} A. Righi, {\it{Electromagnetic fields}}, Nuovo Cimento, 2 (1901) 104

\bibitem{nisbet} A. Nisbet, {\it{Hertzian electromagnetic potentials and associated gauge transformations}}, Proc. R. Soc. Lond. A 231 (1955) 250

\bibitem{jackson} J. D. Jackson, {\it{Classical electrodynamics}}, John Wiley $\&$ Sons (1999) 

\bibitem{DeTe} S. Deser and C. Teitelboim, {\it{Duality Transformations of Abelian and Nonabelian Gauge Fields}}, Phys. Rev. D 13 (1976) 1592

\bibitem{as} G. N. Afanasiev and Yu. P. Stepanovsky, {\it{The helicity of the free electromagnetic field and its physical meaning}}, Nuovo Cimento A 109 (1996) 3

\bibitem{bliokhnjp} K. Y. Bliokh, A. Y. Bekshaev and F. Nori, {\it{Dual electromagnetism: helicity, spin, momentum and angular momentum}}, New J. Phys. 15 (2013) 033026

\bibitem{cameronnjp} R. P. Cameron and S. M. Barnett, {\it{Electric-magnetic symmetry and Noether's theorem}},  New J. Phys. 14 (2012) 123019

\bibitem{EDHZ-heli} M. Elbistan, C. Duval, P. A. Horvathy and P.-M. Zhang,
 {\it{Duality and helicity: a symplectic viewpoint}},
 Phys. Lett. B 761 (2016) 265 

\bibitem{CMYnjp14} R. P. Cameron, S. M. Barnett and A. M. Yao, {\it{Optical helicity, optical spin and related quantities in electromagnetic theory}}, New. J. Phys. 14 (2012) 053050

\bibitem{BBr2} I. Bialynicki-Birula and Z. Bialynicka-Birula, {\it{Quantum-mechanical description of optical beams}}, J. Opt. 19 (2017) 125201

\bibitem{hb} M. Y. Han and L. C. Biedenharn, {\it{Manifest duality invariance in electrodynamics and the Cabibbo-Ferrari theory of magnetic monopoles}}, Nuovo Cimento 2A (1971) 544

\bibitem{BBr1} I. Bialynicki-Birula and Z. Bialynicka-Birula, {\it{The role of the Riemann-Silberstein vector in classical and quantum theories of electromagnetism}, J. Phys. A: Math. Theor. 46 (2013) 053001}

\bibitem{bliokhst} J. Dressel, K. Y. Bliokh and F. Nori, {\it{Spacetime algebra as a powerful tool for electromagnetism}}, Phys. Rep. 589 (2015) 1

\bibitem{fcorbaton}
 I.~Fernandez-Corbaton, et al., {\it{Electromagnetic duality dymmetry and helicity Conservation for the Macroscopic Maxwell's Equations}}, Phys. Rev. Lett. 111 (2013) 060401

\bibitem{bliokhm} K. Y. Bliokh, Y. S. Kivshar and F. Nori, {\it{Magnetoelectric effects in local light-matter interactions}}, Phys. Rev. Lett. 113 (2014) 033601

\bibitem{bliokhcm} F. Alpeggiani, K. Y. Bliokh, F. Nori and L. Kuipers, {\it{Electromagnetic helicity in complex media}}, arXiv: 1802.09392

\bibitem{zb} A.A. Zyuzin and A. A. Burkov, {\it{Topological response in Weyl semimetals and the chiral anomaly}}, Phys. Rev. B 86 (2012) 115133

\bibitem{qhz} X-L. Qi, T. L. Hughes and S-C. Zhang, {\it{Topological field theory of time-reversal invariant insulators}}, Phys. Rev. B 78 (2008) 195424

\bibitem{ranada2} A. F. Ra\~{n}ada, {\it{On the magnetic helicity,}} Eur. J. Phys. 13 (1992) 70

\bibitem{knotsrev} M. Array\'as, D. Bouwmeester and J. L. Trueba, {\it{Knots in electromagnetism}}, Phys. Rep. 667 (2017) 1

\bibitem{giosha} M. Giovannini and M. E. Shaposnikov, {\it{Primordial hypermagnetic fields and the triangle anomaly}}, Phys. Rev. D 57 (1998) 2186

\bibitem{rcam} R. Cameron, {\it{On the 'second potential' in electrodynamics}}, J. Opt. 16 (2014) 015708

\bibitem{eweyl} M. Elbistan, {\it{Weyl semimetal and topological numbers}}, Int. J. Mod. Phys. B 31 (2017) 1750221

\bibitem{wilczek} F. Wilczek, {\it{Two applications of axion electrodynamics}}, Phys. Rev. Lett. 58 (1987) 1799 

\bibitem{hky} Y. Hirono, D. E. Kharzeev and Y. Yin, {\it{Self-similar inverse cascade of magnetic helicity driven by chiral anomaly}}, Phys. Rev. D 92 (2015) 125031 

\bibitem{bnp} M. Baggio, V. Niarchos and K. Papadodimas, {\it{Aspects of Berry phase in QFT}}, JHEP 04 (2017) 062

\bibitem{yamaxion} S. Ozaki and N. Yamamoto, {\it{Axion Crystals}}, JHEP 08 (2017) 098

\bibitem{bcy} S. M. Barnett, R. P. Cameron and A. M. Yao, {\it{Duplex symmetry and its relation to the conservation of optical helicity}}, Phys. Rev. A 86 (2012) 013845

\bibitem{EHZdual2} M. Elbistan, P. A. Horvathy and P.-M. Zhang, {\it{Duality and helicity: the photon wave function approach}},  Phys. Lett. A 381 (2017) 2375

\bibitem{BBrev} I.~Bialynicki-Birula, {\it{Photon wave function}}, Prog. Opt. 36 (1996) 245  




































\end{thebibliography}
\end{document}